# EV Charging Optimization based on Day-ahead Pricing Incorporating Consumer Behavior

Qun Zhang, Gururaghav Raman, and Jimmy Chih-Hsien Peng
Department of Electrical and Computer Engineering
National University of Singapore, Singapore-117583
Email: e0269754@u.nus.edu

*Abstract*—With the increasing penetration of electric vehicles (EVs) into the automotive market, the electricity peak demand would increase significantly due to home-EV-charging. This paper tackles this problem by defining an 'ideal' EV consumption profile, from which a day-ahead pricing model is derived. Based on historical residential EV-use data ranging over a year, we demonstrate that the proposed optimization process results in a pricing profile that achieves a dual objective of minimizing the total electricity cost, as well as the peak aggregate system demand. Importantly, the proposed formulation is simple, and accounts for the tradeoff between consumer convenience in terms of the number of available charging slots during a day and the reduction in the total electricity cost. This technique is demonstrated to be scalable with respect to the size of the community whose EV charging demands are being optimized.

## I. Introduction

The automotive sector is a major source of carbon emissions. The main emission from this sector, $CO_2$, is widely regarded as an anthropogenic greenhouse gas, affecting the climate [1]. In 2016, the transportation sector in the US overtook electricity generation in terms of the total metric tons of $CO_2$ emitted into the atmosphere [2], which suggests that $CO_2$ produced by traditional car-fuel consumption has had a significant impact on climate change. As an alternative, EVs are experiencing higher penetrations in the market in order to halt the increasingly negative effects of fossil fuel-fed transportation. The energy consumed by EVs could be derived from renewable sources of energy such as solar photovoltaic, wind and biomass generation. Besides, EVs also offer other benefits such as lower operational costs [3] [4].

However, the grid must transform so as to keep up with the demand for electrified transportation [5]. With the development of more efficient and cheaper charging hardware, ever increasing public charge stations, and reducing costs, a charging point in every home seems to be achievable so that EVs can be charged whenever required. Under a new draft directive of Europe Union in October 2016, it would be mandatory for every new or refurbished home in Europe to have an electric car charging point by 2019, in a move that has been considered vital to the success of EV uptake. However, a potential problem for charging EV at home is that the peak of total power demand for a household moves up compared with the residential power demand alone. Residents' electricity is usually concentrated at 18:00-20:00 [6], the same period during which residents would charge their EVs immediately after parking at home. As a result, at the community scale, the peak demand rises significantly with more EVs. This increase in the system peak demand has become a major technical constraint for the widespread use of electric vehicles.

To tackle this problem, a novel pricing scheme is proposed in this paper to reduce the system peak demand by encouraging customers to avoid charging at peak hours of residential demand through a day-ahead scheduling process. This plan requires two prerequisites. One is hardware support such as the presence of smart meter at each home, which can be connected to an electric vehicle controlling charging switch according to the programmed scheduling [7]. The other assumption is that the residential electricity consumption will not be affected by the electricity price, meaning that lifestyle is not changed by electricity prices, with the purpose of measuring the effect of price on EV charging manners. Alternatively, this constraint may be justified by assuming that variable pricing only exists for EV charging, and not for the conventional home appliance demand. Literature survey in EV charging indicates that most previous studies only consider fixed available charging times. For instance, in [8], an EV is supposed to be connected to the grid at 6:15PM, and is scheduled to leave at 7:30AM on the next day. Clearly, there is lack consideration of the user's preferences about when to charge before the scheduling process is done. To the contrary, in this work, a randomized algorithm is adopted to choose the available timeslots with respect to the users' charging probabilities.

Our contributions are summarized as follows: (1) we develop a day-ahead pricing model based on the difference between the existing demand curve and desired 'ideal' demand curve in order to minimize the peak power demand, and (2) we propose a novel method to choose the available charging timeslots according to consumer charging behavior, exploring the balance point of electricity cost and daily convenience from the user's perspective. Overall, for a realistic test system, we demonstrate a reduction in the peak demand of 25.93%, via EV charging scheduling with mixed-integer linear programming (MILP) based on the price, available charging times, as well as the physical constraints of EV-charging.

## II. Proposed Day-Ahead Pricing Scheme

The proposed technique for EV-charging optimization is split into several stages, as shown in Fig. 1. These are explained below.

### A. Ideal electricity consumption profile to minimize the aggregate peak demand

The historical data that is used in this work is obtained from

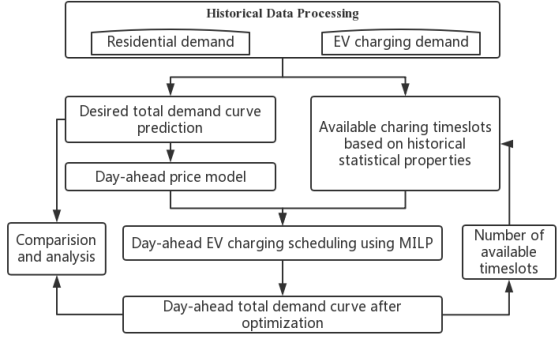

Fig. 1. Flowchart depicting the proposed EV-charging optimization process.

[6], and consists of a set of consumers' residential demands, and their corresponding EV charging schedules. This data is split into weekdays and weekends, as typically, living habit on weekdays and the weekends for most households are very different. To determine the peak demand, the total power demand of the $N$ households in the community is obtained with a day cycle which is divided into $S$ timeslots.

$$X_t = \sum_{i=1}^{N} x_{i,t}, \quad (1)$$

$$Y_t = \sum_{j=1}^{M} y_{j,t}, \quad (2)$$

$$H = \arg(\max(X_t)). \quad (3)$$

Here, $x_{i,t}$, $y_{j,t}$ are respectively the residential demand, and EV charging power at the $t_{th}$ timeslot for $i_{th}$ household and $j_{th}$ EV. $X_t$ is a vector of total residential demand for households during one day, and $Y_t$ is the vector of total charging power. $H$ is the index of timeslot with peak residential demand.

The 'ideal' demand curve with the minimum possible peak demand is determined while keeping the total power demand constant (see Fig. 2 and 3). This desired total demand curve is based on the total electricity demand of all the households over one year, for the purpose of minimizing the peak demand and smoothing the curve. This means that the EV charging power would be preferably allocated to the timeslots with a smaller residential demand. This is represented by the following optimization problem:

$$\text{Minimize } s_1 + s_2 \quad (4)$$

$$s.t. \quad \frac{1}{H-s_1}\sum_{t=1}^{H-s_1}(X_t + E_t) -$$
$$\frac{1}{S-(H-s_2)}\sum_{t=H-s_2+1}^{S}(X_t + E_t) = 0, \quad (5)$$

$$\sum_{t=1}^{H-s_1} E_t + \sum_{t=H-s_2+1}^{S} E_t = \sum_{t=1}^{S} Y_t. \quad (6)$$

Here, $s_1$ and $s_2$ are number of timeslots before and after peak demand point, and $E_t$ the desired average EV charging demand, the sum of which is equal to that obtained from the real historical data. The total demand of the residential and EV powers of the two sides should be the same in order to smooth the curve to the maximum extent. The objective here is to minimize the length of timeslots with the highest power demand. Constraint (5) aims to construct straight lines before $s_1$ and after $s_2$ with the same height, and constraint (6) is set to satisfy that the sum of EV charging power before $s_1$ and after $s_2$ equal to total original EV charging power, referring to Fig. 2 and Fig. 3 in the simulation part for clarity.

### B. Day-ahead pricing model to achieve the ideal profile of consumption

A day-ahead pricing model [9] is presented to solve the problem of increasing peak demand. The concession of comfort preference for consumers to charge EVs is attractive because charging is used to store energy in advance instead of real-time use as other appliances. EV charging can be turned on or off with intelligent-based control of smart meters without violating the minimum off-on interval. Therefore, the price is an incentive strategy for consumers to modify their charging time, while maintaining a balance between the cost and convenience. We would like to determine how much concession a consumer would make when the price is for the curve nearest to the ideal profile as explained in Section II. A:

$$f_t = f_{min} + (f_{max} - f_{min})\frac{E_t - E_{min}}{E_{max} - E_{min}}, \quad (7)$$

where $f_t$ is the price, and $f_{min}$ and $f_{max}$ are its lower and upper bounds. Normalization is proposed to compress this curve into a practical price range, assuming that the previous constant price is the average of the bounds. $E_{min}$ and $E_{max}$ are the minimum and maximum values of $E_t$ over all the time slots.

### C. Modeling randomly-available timeslots for EV charging

*1) Statistical EV charging behavior:* Users' habits regarding EV charging can be gleaned from the historical data. For each household, the times of charging at each timeslot are accumulated and then divided by the number of days; the result is regarded as the frequency of charging (this reflects the consumer's charging behavior). In other words, if the charging frequency of an EV at the $i_{th}$ timeslot is the biggest, the probability of this EV to be charged at this time is assumed to be the highest.

*2) Available charging timeslots:* The historical data recorded the charging behaviors of the users. However, the number of available charging timeslots should be determined accurately so that optimization could be performed to minimize the electricity cost. Here we define a parameter called Comfort Tolerance ($\tau$), which measures how much concession customers would give to comfort in order to achieve the least electricity cost that would result in the ideal demand curve:

$$\tau = \frac{T_i^{avai}}{S}. \quad (8)$$

$T_i^{avai}$ is the number of chosen available charging durations of $EV_i$ during a day. Thus, a higher value of $\tau$ reflects that the

customer would compromise more comfort in exchange for monetary benefit.

Once the total number of available charging timeslots is determined, EV users are selected randomly in the order of descending charging probability for each household. This means that timeslot allocations are not fixed; there is an inherent randomness in the allocation process. For this, a roulette wheel selection scheme is proposed. A turntable is divided into $S$ timeslots, and the area of each EV is proportional to its charging frequency, and a random user is selected from here. This process is repeated until all the available timeslots are filled.

*D. Optimization problem formulation*

*1) Optimization model:* The objective of EV charging scheduling is to determine the optimal charging time and power such that the electricity cost is minimal, while at the same time being constrained in terms of the charging level, storage capacity, and consumers' convenience. The objective function is expressed as follows:

$$minimize\ C_i = \sum_{t=1}^{T_i}\{P_{i,t}\cdot \Delta t \cdot f_t\}, \quad (9)$$

where $C_i$ is electricity cost during the scheduled time of $EV_i$, $T_i$ the length of the charging period for $EV_i$, $P_{i,t}$ the power consumed by $EV_i$ at the $t_{\text{th}}$ timeslot, $\Delta t$ the duration for each timeslot, and $f_t$ the electricity price during the $t_{\text{th}}$ timeslot.

The following constraints are imposed on the above optimization objective:

- *Allowable charging time*: An EV is only allowed to be charged during available timeslots defined as a vector $T_i$ with the length of $T_i^{avai}$.

$$P_{i,t} = 0 \quad \forall t \notin T_i, \quad (10)$$

- *Electric energy balance for interruptible loads*: EV is regarded as an interruptible load [10] which is capable to charge continuously within the bound of charging power (i.e., $P^{min}$ and $P^{max}$). Given the above condition, charging status of EV throughout the scheduling horizon should satisfy:

$$\sum_{t\in T_i} P_{i,t}\cdot \Delta t = E_i, \quad (11)$$

$$P^{min} \le P_{i,t} \le P^{max}, \quad (12)$$

- *Frequent switching limit*: In order to extend the lifespan of the electrical machinery, off-on intervals should be limited as follows:

$$Y_{i,t-1} - Y_{i,t} + Y_{i,k} \le 1$$
$$\forall K: 1 \le K \le T_{off-on}, K \in N^+$$
$$\forall t \in (1, N - T_{off-on} + 1), t \in N^+. \quad (13)$$

where $T_{off-on}$ represents the minimum off-on interval for EV charging, and $Y_{i,t}$ is defined as the status of mechanical part of EV ("1" for "on" and "0" for "off").

*2) MILP solver:* It can be seen that the formulated optimization problem is a MILP [11], as variables in (13) are limited to integers, while in the other two constraints, no such restriction exists. Different from LP, the time complexity for MILP formulations is not linear or polynomial, and these form NP-hard problems. Improvements such as resolving, generating cutting planes, and applying primal heuristics could be utilized to improve the efficiency of the solver. There are some popular optimization solvers to solve MILP such as CPLEX, GUROBI, MOSEK and XPRESS. Here we use GUROBI, which is a commercial solver with high efficiency and supports a variety of programming and modeling languages.

III. CASE STUDY AND SIMULATION RESULTS

*A. Assessment metrics*

To assess the effectiveness of the proposed price-determining algorithm as well as EV charging scheduling compared with original profiles, we use the following metrics:

*1) Peak demand:* The power distribution grid is designed to deal with the highest possible peak demand. However, if this limit is exceeded during its operation, blackouts could result. The objective of this paper is to reduce the peak of aggregate domestic and EV charging demands.

*2) Ramp rate:* The ramp rate is the time-rate of change of the instantaneous real power consumption in the system. Limits are established on this to prevent undesirable effects due to rapid changes in loading or discharge, including large frequency deviations and the loss of stability. Ideally, the ramp rates have to be as minimal as possible.

*3) Variance of the demand curve:* Lower fluctuations in the demand reflects a better stability of electricity consumption curve. As is known, the ideal curve is shaped like a combination of straight line and smooth curve minimizing the variance. Thus, smaller variance of the curve is expected, representing the flat curve.

We mention here that the objective of the proposed price-curve determination algorithm is used to minimize the peak demand. However, the ramp rate and variance are used to determine the quality of the resultant demand profile.

*B. Load profiles before EV-schedule optimization*

In order to test the proposed EV scheduling model, we refer to the 2009 RECS data set for the Midwest region of the US. This includes the historical demand profiles for 200 households, as well as in-home EV-charging behaviors for 256 EVs present in these households, assuming Level 1 (1.92 kW) residential charging infrastructure. Using models presented in [6], realistic load profiles are simulated, which reproduce realistic residential power consumption patterns and EV charging profiles. We assume that all of these households belong to a common community, and the data represent the net system demand. A day is divided into 144 timeslots, each with a duration of 10 minutes, starting at 12:00AM (midnight). The average electricity usage for residential and (residential + EV charging) profiles from this dataset is depicted in Fig. 2.

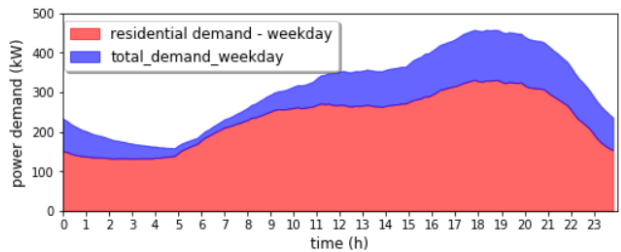

Fig. 2. Original residential-only and total demands of the community.

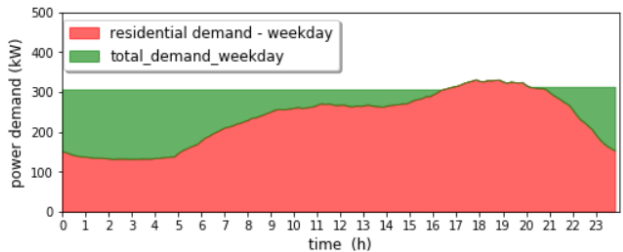

Fig. 3. 'Ideal' power demand of the community.

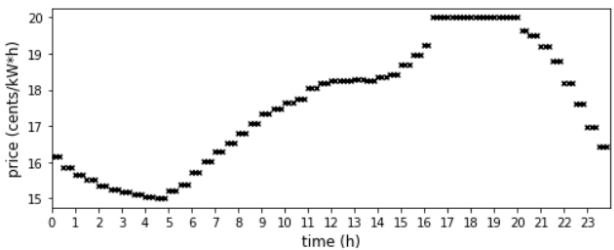

Fig. 4. The desired day-ahead price curve.

From the data above, the ideal electricity demand curve is designed, and shown in Fig. 3. From this figure, we see that the peak demand of the residential curve is high during weekdays. If the EV charging were to happen completely at periods away from the peak period (the desired EV charging behavior of the community), the peak demand of the total system would now equal the peak residential demand.

From the difference of the ideal and original curves, the day-ahead price curve is obtained using (7). Finally, to make the price curve reasonable and close to practical values, lower and upper bounds are set to be $15\cancel{c}/kWh$ and $20\cancel{c}/kWh$. As shown in Fig. 4, the lower price corresponds to a bigger difference of the two curves with the aim to encourage customers charging more during this period. Thus, EV-charging optimization is introduced based on this price profile, the results of which are explained in the sequel.

### C. Simulation results after EV-schedule optimization

For the original test community (200 households), day-ahead EV scheduling is adopted with the assumption that all households are enrolled into the scheme. The simulation parameters are as follows: $P^{max}$ = 1.92 kW, $P^{min}$ = 0.96 kW, and $T_{off-on}$ = 30 min.

As mentioned previously, the available charging time are chosen randomly based on their charging probability. To account for the randomness, each EV optimization is processed for 20 times. The boxplots for the total residential and EV-charging demands after optimization are obtained, and presented in Fig. 6.

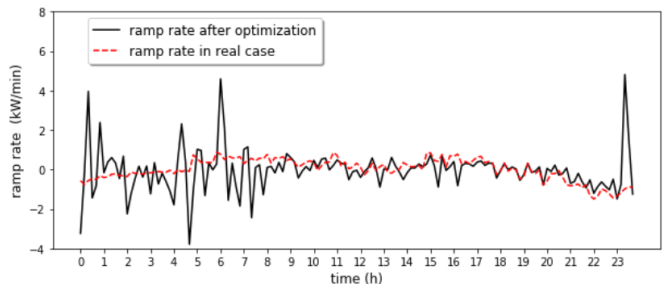

Fig. 5. Ramp rate of the total electricity consumption.

TABLE I
EFFECTIVENESS OF OPTIMIZATION-REDUCTION OF SYSTEM PEAK DEMAND

| Size | Number of timeslots | Original peak demand (kW) | Ideal peak demand (kW) | Optimized peak demand (kW) | Reduction (%) |
|---|---|---|---|---|---|
| 1 | 129 | 457.988 | 331.621 | 339.249 | 25.93 |
| 0.9 | 118 | 408.737 | 294.294 | 337.417 | 17.40 |
| 0.8 | 107 | 366.251 | 267.234 | 320.816 | 12.41 |
| 0.7 | 98 | 311.441 | 225.748 | 281.271 | 9.68 |
| 0.6 | 88 | 259.742 | 192.400 | 240.983 | 7.22 |
| 0.5 | 78 | 222.096 | 161.601 | 198.267 | 10.72 |

From this figure, it is clear that the system demand is close to the ideal profile, with a significant reduction in the aggregate peak demand. Again, this result is for 200 households; when the size of the community increases, peak demand for the whole community can be reduced to a larger extent, as more and more EVs are available for optimization. The corresponding ramp rate is presented in Fig. 5, with shows that ascent or descent rate is within an acceptable range of [-4kW/min, 5kW/min], about 1% of the total demand. From the perspective of electricity servers, it is tolerable but not ideal.

*1) Scaling the proposed optimization algorithm:* To test the scalability of the scheme for various sizes of the community, different households are randomly chosen to be a part of it, and the metrics are calculated for each case. As shown in Table I, with decreasing community size, the peak demand optimization increases. In the case when only 50% of original community (i.e. 100 households) are regarded as a new community, the peak demand after optimization is close to the original peak demand, suggesting that this scheme is more effective in a large community. One possible reason is that the diversity of EV charging times in a larger community promises available charging times staggered over different periods rather than clustered close together. Although all EVs are under the common pricing scheme, different available charging times restrict the optimization process, avoiding the situation that the peak demand from the rush hours is just shifted to timeslots with the lowest price, which is deemed as excessive optimization. At this point, we would like to note the implications of these results in terms of reducing local demand overloads as well. Even if we reduce the number of houses considered for optimization, we find that the net peak demand is reduced due to the staggering of the EV use patterns, which would avoid overloading of the system feeders.

*2) Impact of the number of users participating in the optimization process:* In reality, not all of users would like to compromise their convenience for a smaller electricity cost. We now keep the community size as the original value of 200 households, and test the flexibility of the proposed scheme considering that only some of consumers are now enrolled in the optimization scheme. We attempt to determine the lower limit of this consumer-group size that would result in the optimized curve not surpassing the original peak demand. Referring to Fig. 7, where consumers participating in the scheme charge their vehicles at the lowest-priced timeslots, we see that as the number of consumers participating declines, so does the reduction of the peak demand.

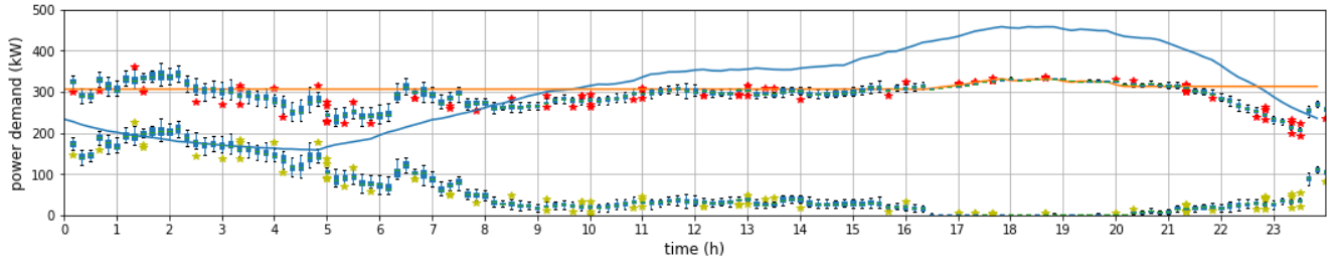

Fig. 6. Electricity demand after optimization of the community with 200 households. The blue and orange curves respectively denote the total system demand before optimization, and the ideal (desired) system demand. The box plots with the green and red outliers respectively denote the EV and total system demands after the proposed optimization process, obtained as a result of 20 trials.

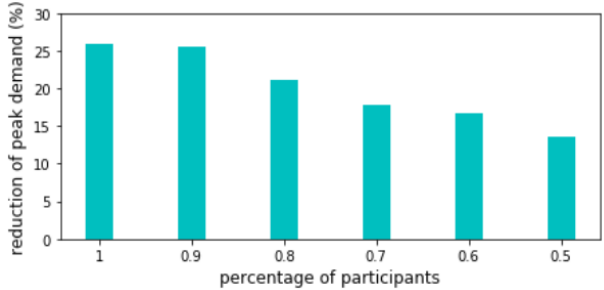

Fig. 7. Peak demand reduction for different percentages of participants.

TABLE II
ELECTRICITY COST SAVING IN EV CHARGING AFTER OPTIMIZATION

| EV No. | Charging demand (kWh) | Cost before optimization ($/kWh) | Cost after optimization ($/kWh) | Cost savings |
|---|---|---|---|---|
| 1 | 5.04 | 0.83 | 0.77 | 7.23% |
| 2 | 4.53 | 1.12 | 0.68 | 39.2% |
| 3 | 6.13 | 1.53 | 0.93 | 93.3% |
| 4 | 8.33 | 1.55 | 1.30 | 16.1% |
| 5 | 8.48 | 1.61 | 1.32 | 18.0% |

To illustrate the cost savings brought about by the proposed optimization process, five households are chosen randomly, and the difference of their electricity costs before and after optimization are presented in Table II. It is assumed here that the price curve is employed for all of the consumers no matter whether they join the scheme or not. Clearly, all consumers save on electricity costs, but to varying degrees depending on their original charging behavior.

## IV. CONCLUSION

A novel day-ahead electric vehicle charging scheduling scheme in a community scale is proposed to tackle the challenge of increased peak demand caused by popularity of EVs and introduction of home-charging infrastructure. The developed electricity price model is based on the current demand curve of the community, and serves the dual objective of reducing the total peak demand, and at the same time minimizing the EV-charging electricity costs. Importantly, one of the contributions of this study is in considering the tradeoff between consumer comfort and the achieved cost savings. Central to this is the random allocation of the available EV charging slots to the various users depending on the probability of use of the respective vehicles. The scalability and flexibility of the proposed scheme are demonstrated using numerical simulations for varying community sizes, and the percentage of residents that enroll into the scheduling program.